\documentclass[aps,prl,superscriptaddress,longbibliography,twocolumn,10pt]{revtex4-2}

\usepackage{graphicx}
\usepackage{amsmath,amssymb,mathtools,bbm,bbold,braket,hyperref,xcolor,multirow,setspace,dsfont,microtype}
\usepackage[utf8]{inputenc}
\usepackage[english]{babel}
\usepackage{pgf}
\hypersetup{colorlinks=true, linkcolor=blue, citecolor=blue, urlcolor=blue}
\usepackage{comment}
\usepackage{graphicx}   % <— handles \includegraphics
\usepackage{mwe}
\usepackage{enumitem}

\hypersetup{colorlinks=true, linkcolor=blue, citecolor=blue, urlcolor=blue}
\definecolor{maroon}{cmyk}{0,0.87,0.68,0.32}

\newcommand{\bea}{\begin{eqnarray}}
\newcommand{\eea}{\end{eqnarray}}

\usepackage{pgf}
\usepackage{layouts}
\usepackage{booktabs} % For elegant tables
\usepackage{array}    % For custom column widths

\usepackage{etoolbox}
\usepackage{longtable}%
\usepackage{colortbl}%
  
\usepackage{booktabs}

\begin{document}

\title{{Scaling of Superradiant Peak Emission in Spatially Extended Emitter Arrays}}

\author{Raphael Holzinger}
\email{raphael$_$holzinger$@$fas.harvard.edu}
\affiliation{Department of Physics, Harvard University, Cambridge, Massachusetts 02138, USA}

\author{Susanne F. Yelin}
\email{syelin$@$g.harvard.edu}
\affiliation{Department of Physics, Harvard University, Cambridge, Massachusetts 02138, USA}

\begin{abstract}
In quantum optics, superradiance is a phenomenon in which a system of $N$ fully excited quantum emitters radiate intense flashes of light during collective decay. However, computing its peak intensity exactly for many spatially separated emitters remains challenging due to the exponential growth of the underlying Hilbert space with system size $N$. Based on third-order cumulant expansion methods, we present general scaling laws for the expononent of the peak emission rate as a function of the emitter number in free-space emitter arrays and arrays coupled to one-dimensional waveguide reservoirs.
We find, that for 1D chains in free-space the peak emission rate scales linearly with $N$, while for 2D and 3D arrays with finite emitter spacing it scales superlinearly but sub-quadratically. For emitter chains coupled to waveguide reservoirs we find that the peak emission rate scales quadratically with $N$.
\end{abstract}

\maketitle
%%%%%%%%%%%%%%%%%%%%%%%%%%%%%%%%%%%%%%%%%%%%%%%%%%%%%%%%%%%%%%%%%%%%%
\textbf{Introduction} -- Originally introduced by Dicke~\cite{Dicke_originalpaper}, superradiance describes the drastic enhancement of collective emission from an ensemble of $N$ dipole emitters coupled to a common electromagnetic environment. When the system is initialized in the {fully excited} state, spontaneous phase synchronization among emitters leads to a transient burst of radiation with intensity far exceeding that of uncorrelated decay. In Dicke’s idealized model, all emitters are indistinguishable and positioned at the same point, resulting in a peak emission rate that scales quadratically with emitter number, $R_\mathrm{peak} \!\propto\! N^2\Gamma$~\cite{gross_haroche,lemberger2021radiation,malz2022large}. This situation, often referred to as \emph{Dicke superradiance}~\cite{gross_haroche,Anatolii_V_Andreev1980-py,haake1972quantum}, is a cornerstone of cooperative quantum optics~\cite{koppenhofer2022dissipative,norcia2016superradiance,ferioli2022observation,Reitz2022}, and admits compact analytical solutions~\cite{holzinger2025solvingdickesuperradianceanalytically,holzinger2025exactanalyticalsolutiondicke}.

\begin{figure}[ht]
    \centering   
\includegraphics[width=0.85\columnwidth]{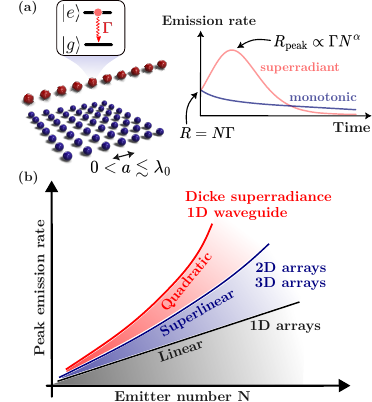}
\caption{(a) Schematic of a spatially extended ensemble of $N$ two-level quantum emitters with spontaneous decay rate $\Gamma$ and finite nearest-neighbor separation $a > 0$. When the fully excited state $|e\rangle^{\otimes N}$ undergoes superradiant decay, the photon emission rate $R$ (in photons per second) reaches a pronounced peak $R_\mathrm{peak}$ ($\alpha \!> \!1$), departing from monotonic or exponential decay ($\alpha \!= \!1$). (b) In this work we study spatially extended arrays in free-space or coupled to bidirectional waveguide reservoirs based on a third-order cumulant expansion method.}
\label{fig1}
\end{figure}

While the Dicke limit is never strictly realized in free space, it can be effectively engineered in cavity and waveguide QED platforms, where emitters couple uniformly to a single or few photonic modes~\cite{goban2015superradiance,PhysRevX.14.011020,sheremet2023waveguide,Solano2017}. By contrast, most experimentally relevant systems are \emph{spatially extended ensembles}, where emitter positions are distinguishable and dipole-dipole couplings depend on their separation. Examples include cold atomic gases~\cite{gross1976observation,PhysRevA.75.033802,PhysRevA.95.043818,PhysRevLett.51.1175,PhysRevLett.117.073002,ferioli2021laser,ferioli2022observation}, solid-state emitters~\cite{Scheibner2007,Rainò2018,Bradac2017,https://doi.org/10.1002/adfm.202102196,Jahnke2016,Cong_2016}, molecular systems~\cite{lange2024superradiant,Kim2023}, and nuclear ensembles~\cite{Chumakov2018}. In this setting, a central question is how the {peak} photon emission rate $R_\mathrm{peak}$ scales with system size and geometry, and whether any form of superlinear scaling with $N$ persists once spatial structure is taken into account (see Fig.~\ref{fig1}).

The main theoretical challenge arises from the fact that spatially distinguishable emitters span a Hilbert space that grows exponentially with $N$. Exact treatments of the full master equation are therefore restricted to $N \lesssim 20$~\cite{CARMICHAEL2000417,Molmer:93,shammah2018open}. Approximate numerical techniques such as mean-field methods, cluster expansions, phase-spase methods~\cite{Mink_2023} or few-order cumulant truncations allow one to go beyond mesoscopic sizes of $N\sim 10^2$~\cite{rubies2023characterizing,masson2022universality,Robicheaux_cumulants}.
%%%%%%%%%%%%%%%%%%%%%%%%
%%%%%%%%%%%%%%%%%%%%%%%%%%%%%%%%%%%%%%%%%%%%%%%%%%%%%%%%%%%%%%%%%%%%%
At the same time, recent theoretical advances have clarified several fundamental aspects of superradiance. Universal upper bounds on the maximal peak emission rate in atomic arrays prepared in pure states suggest that superlinear scaling with $N$ is \emph{in principle} possible beyond the Dicke limit~\cite{mok2025universalscalinglawscorrelated}. Whether such behavior is actually realized for the experimentally relevant situation of a fully inverted ensemble undergoing collective decay has remained open. This scenario, in which all emitters are initially excited and no external drive is present, has been the focus of theoretical work for over 70 years~\cite{Dicke_originalpaper,bonifacio1971quantum,gross_haroche,AGARWAL19731,Super_theory_Rehler,shammah2018open,masson2022universality,zhang2025unravelingsuperradianceentanglementmutual,rosario2025unravelingdickesuperradiantdecay}. In particular, Refs.~\cite{robicheaux2021theoretical,masson2022universality} provided elegant criteria for the \emph{existence} of a superradiant peak in extended systems, but did not yield quantitative values or scaling laws for $R_\mathrm{peak}$ itself.

\begin{figure*}[ht]
    \centering   
\includegraphics[width=0.8\textwidth]{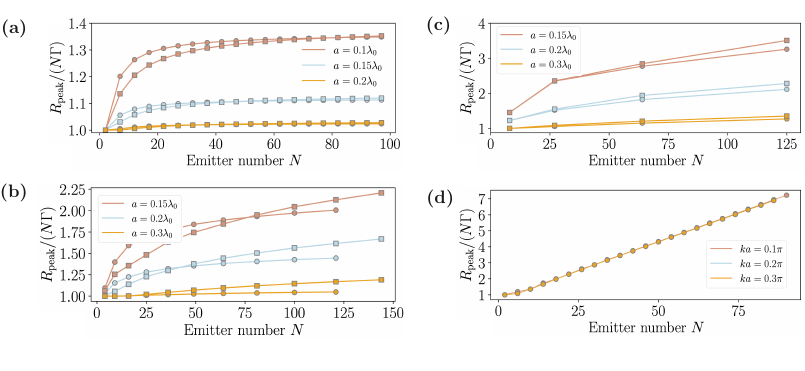}
\vspace{-2em}
\caption{Peak photon emission rate as a function of the emitter number $N$. (a) one-dimensional euqidistant chain, (b) two-dimensional square array (c) three-dimensional cubic array. For the chain we find a linear scaling of the peak emission rate with $N$ while for the square and cubic arrays the peak emission rate scales superlinearly but less than quadratically at finite emitter separations. (d) Equidistant emitter chain coupled to a bidirectional waveguide with relative phases between the emitters, $ka/\pi=0.1,0.2,0.3$. We find, that $R_\mathrm{peak}\propto \Gamma N^2$, irrespective of the relative phase. In (a) for nearest-neighbor spacings $a/\lambda_0=0.1,0.15,0.2$, in (b-c) $a/\lambda_0=0.15,0.2,0.3$. Circles ($\circ$) correspond to circular $(1,\pm i,0)/\sqrt{2}$ and squares ($\square$) to linear polarized $(0,0,1)$ dipoles. For all plots a third-order cumulant expansion has been used.}
\label{fig2}
\end{figure*}

In this work we address this problem by combining a microscopic description of collective decay in extended arrays with a third-order cumulant expansion. This approach allows us to compute the peak emission rate for large arrays of up to hundreds of emitters in one, two, and three dimensions, and to extract scaling laws as a function of both emitter number $N$ and lattice spacing $a$. As summarized in Fig.~\ref{fig2}, we find that in free-space one-dimensional chains the peak emission rate scales \emph{linearly} with emitter number, $R_\mathrm{peak} \propto N$, while in two- and three-dimensional arrays with finite spacing it scales \emph{superlinearly but subquadratically}, with an exponent $1 < \alpha < 2$ in $R_\mathrm{peak} \sim N^\alpha$. By contrast, for emitter chains coupled to a bidirectional one-dimensional waveguide reservoir, the peak emission rate recovers a \emph{quadratic} scaling $R_\mathrm{peak} \propto N^2$, independent of the relative phase $k a$ between neighboring emitters, as was also found in Refs.~\cite{zhang2025unravelingsuperradianceentanglementmutual,lee2025exactmanybodyquantumdynamics}.  

\begin{figure*}[ht]
    \centering   
\includegraphics[width=1\textwidth]{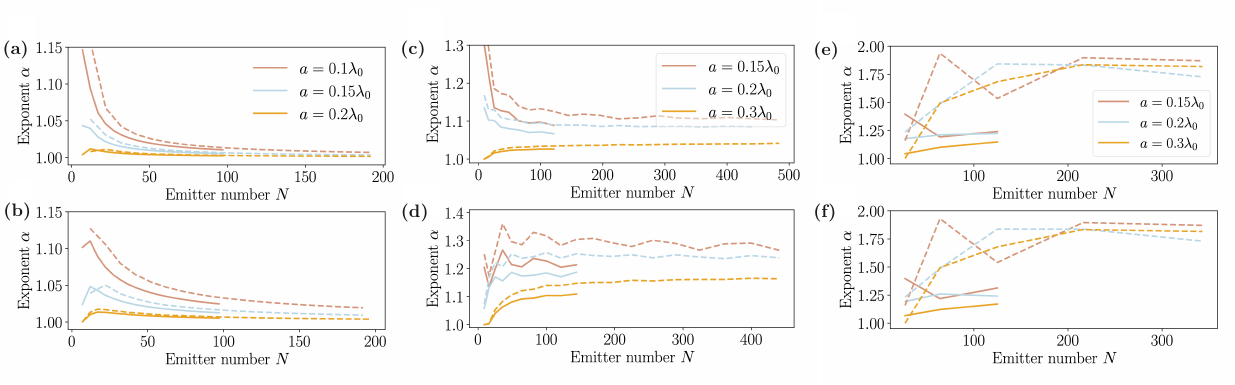}
\vspace{-2em}
\caption{The exponent $\alpha$, assuming the general form for the peak emission rate $R_\mathrm{peak} = \Gamma \beta N^\alpha$ with the same emitter spacings as in Fig.~\ref{fig2}. One-dimensional chain for (a) circular and (b) linear polarized emitters with the exponent $\alpha$ converging to one for increasing $N$. Square array for (c) circular and (d) linear polarization. Cubic array for (e) circular and (f) linear polarization. The continuous lines (-) correspond to a third-order cumulant expansion and dashed lines (- -) to a second-order cumulant expansion.}
\label{fig3}
\end{figure*}

\begin{figure}[ht]
    \centering   
\includegraphics[width=0.7\columnwidth]{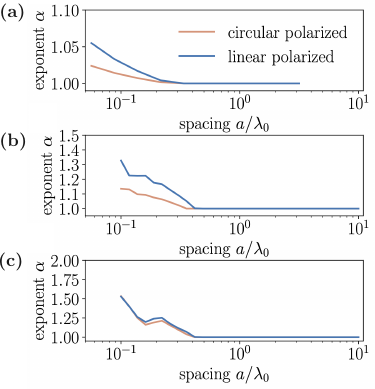}
\caption{The exponent $\alpha$ as a function of emitter spacing $a$ for (a) linear chain, (b) square array and (c) cubic array for linear and circular polarization. The simulations are performed with a third order cumulant expansion. We note, that the exponent will keep decreasing for for increasing emitter numbers at any given spacing.}
\label{fig4}
\end{figure}

To make these scalings precise, we assume a generic scaling form
\begin{equation} \label{emission-formula}
    R_\mathrm{peak}(N,a) = \Gamma\,\beta(a)\,N^{\alpha(N,a)},
\end{equation}
and compute the effective exponent $\alpha(N,a)$ from the numerically obtained $R_\mathrm{peak}$ via logarithmic derivatives. The resulting behavior, shown in Figs.~\ref{fig3} and~\ref{fig4}, reveals that (i) in one-dimensional free-space chains $\alpha(N,a)\to 1$ for large $N$ at any fixed spacing $a>0$, (ii) in 2D and 3D arrays there is a geometry- and spacing-dependent range of $N$ where $\alpha>1$, but $\alpha$ always remains below 2 and eventually returns to unity for large $N$, and (iii) in waveguide QED the exponent approaches $\alpha\simeq 2$ over the entire accessible range. Together, these findings shed some light on when and how superradiant enhancements can be made to scale with system size in spatially extended ensembles.

\textbf{Theoretical description} -- We consider $N$ identical two-level emitters with ground and excited states $\ket{g_n}$ and $\ket{e_n}$, transition frequency $\omega_0 = 2\pi c/\lambda_0$, and individual spontaneous decay rate $\Gamma$. The emitters are located at fixed positions $\{\mathbf{r}_n\}$, which can form one-, two-, or three-dimensional regular arrays with lattice constant $a$, or linear chains coupled to a one-dimensional waveguide (see Fig.~\ref{fig1}). Throughout the dynamics we assume that the emitters are tightly trapped, such that motional effects---center-of-mass motion, recoil, and zero-point fluctuations---can be neglected. In realistic experiments these effects tend to reduce collective couplings and thereby suppress superradiant features; including them would thus quantitatively lower $R_\mathrm{peak}$ and accelerate the convergence toward the linear scalings reported below.

We further assume that all emitters possess the same transition dipole moment $\mathbf{d}$. Changing the dipole orientation modifies the specific values of the dipole-dipole couplings but does not alter the qualitative scaling behavior; in Figs.~\ref{fig2}–\ref{fig4} we explicitly compare linear and circular polarizations.

Tracing out the photonic reservoir within the standard Born–Markov and rotating-wave approximations~\cite{charmichael_1,charmichael_2,Lehmberg1970_1}, the emitter density matrix $\rho$ obeys the Lindblad master equation
\begin{equation}
    \dot{\rho} = -\frac{i}{\hbar}[\mathcal{H},\rho] 
    + \sum_{n,m=1}^{N} \Gamma_{nm}
    \left(
        \sigma_n \rho \sigma_m^\dagger
        - \frac{1}{2} \{ \sigma_n^\dagger \sigma_m, \rho \}
    \right).
    \label{master}
\end{equation}
Here $\sigma_n = \ket{g_n}\!\bra{e_n}$ is the lowering operator of the $n$-th emitter, and the Hamiltonian
\begin{equation}
    \mathcal{H} = \sum_{n\neq m} J_{nm}\,\sigma_n^\dagger \sigma_m
\end{equation}
describes coherent exchange of excitations between emitters. The complex coefficients $J_{nm}$ and $\Gamma_{nm} = \Gamma_{mn}$ represent coherent and dissipative dipole-dipole couplings, respectively, and are determined by the electromagnetic Green’s tensor of the environment. For a three-dimensional photonic reservoir (free space), they are given by~\cite{GreensFunction_novotny_hecht_2006} (see Supplemental Material)
\begin{equation}
    J_{nm} - \frac{i}{2}\Gamma_{nm}
    = -\frac{3\pi\Gamma}{\omega_0}\,
       \mathbf{d}^\dagger
       \cdot \mathbf{G}\big(\mathbf{r}_{nm},\omega_0\big)
       \cdot \mathbf{d},
\end{equation}
where $\mathbf{r}_{nm} = \mathbf{r}_n - \mathbf{r}_m$ and $\mathbf{G}$ is the Green’s tensor. The diagonal elements satisfy $\Gamma_{nn} = \Gamma$, while the off-diagonal elements $\Gamma_{n\neq m}$ encode collective, geometry-dependent decay. For emitters coupled to a single-mode bidirectional waveguide, the corresponding expressions simplify to infinite-range couplings with a phase factor $\exp\big(i k_0 |x_n-x_m|\big)$, as discussed in the Supplemental Material.
%%%%%%%%%%%%%%%%%%%%%%%%%%%%%%%%%%%%%%%%%%%%%%%%%%%%%%%%%%%%%%%%%%%%%
%%%%%%%%%%%%%%%%%%%%%%%%%%%%%%%%%%%%%%%%%%%%%%%%%%%%%%%%%%%%%%%%%%%%%
Our interest lies in the radiated photon flux, or emission rate,
\begin{equation}
    R(t) = -\frac{d}{dt}\sum_{n=1}^{N} \langle \sigma_n^{ee} \rangle
         = \sum_{n,m=1}^{N}\Gamma_{nm} 
           \langle \sigma_n^\dagger \sigma_m \rangle,
    \label{emission}
\end{equation}
where $\sigma_n^{ee} = \ket{e_n}\!\bra{e_n}$, and we have used Eq.~\eqref{master} to express the time derivative in terms of two-point correlators. At $t=0$, for a fully excited initial state $\ket{\psi_0} = \ket{e}^{\otimes N}$, the emission rate reduces to the incoherent value $R(0) = N\Gamma$. For sufficiently small lattice spacings $a\ll\lambda_0$, correlations develop rapidly and a superradiant burst with $R_\mathrm{peak} > N\Gamma$ emerges at a finite time $t_\mathrm{peak}>0$.

The role of the Hamiltonian $\mathcal{H}$ in such fully inverted, collectively decaying systems is subtle: it induces coherent excitation exchange, which tends to dephase the collective dipole and thereby suppress superradiance. Previous numerical studies~\cite{masson2022universality,holzinger2025collectivesuperradianceestimatingpeak} have shown that, for the parameters of interest here, including $\mathcal{H}$ leads to only a modest quantitative reduction of $R_\mathrm{peak}$, without affecting the scaling exponents with $N$ and $a$. In line with these findings, and to focus on dissipative scaling, we neglect $\mathcal{H}$ in what follows.

\textbf{Third-order cumulant expansion} -- Direct solution of Eq.~\eqref{master} requires evolving a density matrix in a Hilbert space of dimension $2^N$, which becomes infeasible already for $N\gtrsim 20$. To access much larger systems we work at the level of correlation functions and truncate the resulting hierarchy using a cumulant expansion~\cite{Cumulant_Kubo}.

We introduce the usual Pauli operators,
\begin{equation}
    \sigma_n^- = \sigma_n, \qquad 
    \sigma_n^+ = \sigma_n^\dagger, \qquad
    \sigma_n^z = \ket{e_n}\!\bra{e_n} - \ket{g_n}\!\bra{g_n},
\end{equation}
and derive from Eq.~\eqref{master} the equations of motion for single- and two-body expectation values,
\begin{equation}
    s_n^z = \langle \sigma_n^z \rangle, \qquad
    C_{nm} = \langle \sigma_n^+ \sigma_m^- \rangle \quad (n\neq m),
\end{equation}
which fully determine the emission rate in Eq.~\eqref{emission}. The resulting equations couple to three-body correlators such as $\langle \sigma_n^z \sigma_m^+ \sigma_\ell^- \rangle$, which in turn couple to four-body correlators, and so on, generating an infinite hierarchy.

In a cumulant expansion this hierarchy is truncated by neglecting connected correlations above a given order. Retaining up to third-order cumulants amounts to factorizing expectation values of four operators according to
\begin{align}
    \langle A B C D \rangle
    &\approx 
      \langle A B \rangle \langle C D \rangle
      + \langle A C \rangle \langle B D \rangle
      + \langle A D \rangle \langle B C \rangle
\nonumber\\
    &\quad 
      -2\, \langle A \rangle \langle B \rangle
            \langle C \rangle \langle D \rangle \\
      &+ \text{terms involving lower-order cumulants}, \nonumber
\end{align}
while keeping all connected contributions up to three operators exactly. In practice we evolve $s_n^z$, $C_{nm}$, and the relevant three-body quantities, replacing four-point functions by products of lower-order correlators. A second-order truncation, by contrast, already neglects all three-body connected cumulants and approximates \emph{all} higher-order moments in terms of one- and two-point functions only.

We note, that for regular arrays with identical emitters and equal nearest-neighbor spacings, translational and permutation symmetries dramatically reduce the number of independent variables. In a one-dimensional chain with spacing $a$, for instance, correlators depend only on the separation $|n-m|$, so that the number of distinct $C_{nm}$ scales as $\mathcal{O}(N)$ rather than $\mathcal{O}(N^2)$. Similar reductions occur in two- and three-dimensional square and cubic arrays. Combined with the cumulant truncation, this allows in principle to simulate arrays with several hundreds of emitters while explicitly retaining the build-up of correlations needed to capture superradiant emission.

We compare third- and second-order truncations at larger $N$, as indicated by the solid (third-order) and dashed (second-order) curves in Fig.~\ref{fig3}. For dense arrays, second order systematically underestimates both $R_\mathrm{peak}$ and the effective exponent $\alpha$, demonstrating the necessity of including three-body cumulants to obtain reliable scaling behavior.

%%%%%%%%%%%%%%%%%%%%%%%%%%%%%%%%%%%%%%%%%%%%%%%%%%%%%%%%%%%%%%%%%%%%%
\textbf{Scaling of the peak emission rate} -- Using the third-order cumulant expansion, we now compute the emission dynamics for various geometries and extract the peak emission rate $R_\mathrm{peak}(N,a)$.
%%%%%%%%%%%%%%%%%%%%%%%%
%%%%%%%%%%%%%%%%%%%%%%%%%%%%%%%%%%%%%%%%%%%%%%%%%%%%%%%%%%%%%%%%%%%%%
Figure~\ref{fig2}(a) shows $R_\mathrm{peak}$ as a function of emitter number $N$ for one-dimensional free-space chains with different lattice spacings $a/\lambda_0 = 0.1,0.15,0.2$ and two dipole orientations (linear and circular). For small $N$ we observe a pronounced superradiant enhancement, $R_\mathrm{peak} > N\Gamma$, with the deviation from linearity strongest at the smallest spacing. However, once $N$ becomes sufficiently large, all curves approach a \emph{linear} scaling,
\begin{equation}
    R_\mathrm{peak} \simeq \Gamma\,\beta_{1\mathrm{D}}(a)\,N,
\end{equation}
with an $a$-dependent prefactor $\beta_{1\mathrm{D}}(a)>1$. Importantly, the \emph{exponent} governing the asymptotic dependence on $N$ is unity for all considered spacings, i.e. $\alpha\to 1$ in one-dimensional free-space chains.

In two- and three-dimensional arrays the behavior is richer. Figures~\ref{fig2}(b) and~\ref{fig2}(c) display $R_\mathrm{peak}$ versus $N$ for 2D square and 3D cubic lattices, respectively, again for several spacings $a/\lambda_0$ and both linear and circular polarizations. For small and intermediate system sizes we find a clear \emph{superlinear} scaling, with $R_\mathrm{peak}$ growing faster than $N$, reflecting the larger number of strongly coupled neighbors in higher dimensions. This superlinear regime is particularly pronounced at the smallest spacings considered. In other words, two- and three-dimensional free-space arrays support a wide parameter range where $R_\mathrm{peak}\sim N^\alpha$ with $1<\alpha<2$.

In contrast, Fig.~\ref{fig2}(d) shows that chains of emitters coupled to a bidirectional single-mode waveguide reservoir exhibit a robust \emph{quadratic} scaling,
\begin{equation}
    R_\mathrm{peak} \propto \Gamma N^2,
\end{equation}
for all relative phases $k a/\pi = 0.1,0.2,0.3$ considered~\cite{zhang2025unravelingsuperradianceentanglementmutual,lee2025exactmanybodyquantumdynamics}. This behavior reflects the effectively one-dimensional character of the photonic reservoir: each emitter couples to the same guided mode, so that collective decay rates can grow linearly with $N$, giving rise to $R_\mathrm{peak}\propto N^2$ in close analogy with Dicke superradiance. Remarkably, this quadratic scaling persists even when the array spacing is comparable to the wavelength, highlighting the qualitative difference between one-dimensional and three-dimensional environments.

\textbf{Scaling exponent} -- To quantify these trends more systematically, we extract the exponent $\alpha(N,a)$ in Eq.~(\ref{emission-formula}) from finite differences of $\ln R_\mathrm{peak}$ with respect to $\ln N$:
\begin{equation}
    \alpha(N,a) = \partial_{\ln (N)} \ln (R_\mathrm{peak})
\end{equation}
This procedure allows us to track how the scaling exponent evolves with system size and spacing.

Figure~\ref{fig3} summarizes the resulting exponents for one-, two-, and three-dimensional free-space arrays. Panels~\ref{fig3}(a,b) show $\alpha(N,a)$ for one-dimensional chains with circular and linear polarization, respectively. For all spacings and both polarizations, $\alpha$ starts above unity at small $N$, reflecting the initial growth of cooperative correlations, but then monotonically decreases and approaches $\alpha\simeq 1$ as $N$ increases. The approach is faster for larger spacings, where collective couplings are weaker and the build-up of long-range correlations is limited. The comparison between third-order (solid lines) and second-order (dashed lines) truncations illustrates that neglecting three-body cumulants generally underestimates $\alpha$ in the regime where correlations are strongest.

For two-dimensional square arrays [Figs.~\ref{fig3}(c,d)] and three-dimensional cubic arrays [Figs.~\ref{fig3}(e,f)], the exponent exhibits a pronounced maximum above unity at intermediate $N$, whose height and position both depend on the lattice spacing and the dipole orientation. For the smallest spacings shown, the maximum can significantly exceed $\alpha=1$, indicating a broad regime where the peak emission grows superlinearly with system size. However, as $N$ is further increased, $\alpha(N,a)$ decreases and eventually saturates below two for finite spacings.

A complementary perspective is provided in Fig.~\ref{fig4}, which displays the exponent $\alpha$ as a function of the lattice spacing $a$ for three fixed geometries: (a) linear chains, (b) square arrays, and (c) cubic arrays, again for both linear and circular dipole orientations. For large spacings $a\gg\lambda_0$ the emitters decay essentially independently and $\alpha\to 1$, as expected. As $a$ decreases below the wavelength, collective couplings strengthen and $\alpha$ increases and depends on geometry and polarization. Taken together, Figs.~\ref{fig3} and~\ref{fig4} demonstrate that in free-space arrays superlinear scaling of the peak emission rate is a array size and emitter density dependent.
%%%%%%%%%%%%%%%%%%%%%%%%
%%%%%%%%%%%%%%%%%%%%%%%%%%%%%%%%%%%%%%%%%%%%%%%%%%%%%%%%%%%%%%%%%%%%

\textbf{Conclusions} -- We have investigated the scaling of the peak emission rate in superradiant decay from fully inverted, spatially extended ensembles of quantum emitters. By combining a microscopic master-equation description with a third-order cumulant expansion, we were able to compute the emission dynamics and extract $R_\mathrm{peak}$ for mesoscopic free-space arrays in one, two, and three dimensions, as well as for chains coupled to a bidirectional waveguide reservoir.

Our main findings can be summarized as follows. (i) In one-dimensional free-space chains, the peak emission rate grows linearly with emitter number, $R_\mathrm{peak}\propto N$, for any fixed finite spacing $a>0$, even though substantial superradiant enhancement over the independent-emitter value $N\Gamma$ can occur at small $N$. (ii) In two- and three-dimensional arrays, $R_\mathrm{peak}$ exhibits superlinear scaling $R_\mathrm{peak} \sim N^\alpha$ with $1<\alpha<2$ over an extended range of system sizes and spacings. (iii) In waveguide QED platforms, by contrast, the peak emission rate scales quadratically with emitter number, $R_\mathrm{peak}\propto N^2$, independently of the relative phase $k a$, mirroring the behavior of Dicke superradiance in a single-mode environment.

These results clarify under which conditions superradiant bursts can be made brighter by increasing the number of emitters in spatially extended systems. In three-dimensional reservoirs, the peak emission rate cannot grow faster than linearly with $N$ in the asymptotic limit; any superlinear scaling is necessarily confined to finite system sizes and densities. This establishes fundamental constraints on the achievable brightness of superradiant light sources based on extended atomic or solid-state arrays. At the same time, our analysis identifies geometries and spacings that maximize the superradiant enhancement at a given $N$, providing practical guidelines for the design of optimized superradiant ensembles.

Experimentally, our predictions can be tested in current platforms capable of preparing large, ordered arrays of quantum emitters, such as tweezer arrays of neutral atoms, optical lattices, solid-state emitter arrays, and nanophotonic interfaces. Measuring $R_\mathrm{peak}$ as a function of $N$ and $a$ in these systems would allow one to map out the crossover from superlinear to linear scaling and to determine the optimal emitter number and spacing for a given geometry. In parallel, waveguide QED experiments could probe the predicted quadratic scaling and its robustness to imperfections and disorder. On the theoretical side, extending the present framework to structured and chiral photonic reservoirs~\cite{PhysRevX.14.011020,lodahl2017chiral}, as well as to disordered or partially filled arrays~\cite{rubies2023characterizing}, represents a natural next step. Finally, incorporating directional detection for example, will be important for connecting superradiant scaling laws to experimentally accessible observables~\cite{masson2024dicke,robicheaux2021theoretical}.

\vspace{1em}
\noindent \emph{Acknowledgments -} We thank Avishi Poddar, Ana Asenjo-Garcia and Xin Zhang for valuable discussions. S.F.Y. acknowledge NSF via PHY-
2207972, the CUA PFC PHY-2317134, and QuSeC-TAQS OMA-2326787 in addition to AFOSR FA9550-24-1-0311.

%TC:ignore

%%%%%%%%%%-----------%%%%%%%%%%-----------%%%%%%%%%%-----------%%%%%%%%%% Appendixary Material

\bibliography{apssamp}

\clearpage
\pagebreak
\onecolumngrid

\appendix            % prints the heading only once
\setcounter{equation}{0}          % start from S1
\renewcommand{\theequation}{S\arabic{equation}}

\section{Supplemental Material}

\section{Electromagnetic Green's tensor}\label{appendix:green}
The system consists of $N$ two-level dipole emitters, each with resonance frequency $\omega_0$ and spontaneous decay rate $\Gamma  = \omega_0^3 \mu^2 /(3\pi c^3 \epsilon_0 \hbar )$. By tracing out the electromagnetic field using the Born-Markov approximation~\cite{lalumiere2013input}, the emitter density matrix $\rho$ evolves in time as
\begin{align} \label{master2}
    \dot{\rho} = - \frac{i}{\hbar}[\mathcal{H},\rho] + \sum_{n,m=1}^N \Gamma_{nm} \Big(\sigma_n \rho \sigma_m^\dagger - \frac{1}{2} \{ \sigma^\dagger_n \sigma_m, \rho \}\Big) ,
\end{align}
where $\sigma_n = |g_n\rangle \langle e_n|$ is the spin lowering operator for the $n^{th}$ emitter and the Hamiltonian in the rotating frame of the emitter frequency $\omega_0$ is given by
\begin{equation}
    \mathcal{H} =  \sum_{n,m\neq n}^N J_{nm} \sigma^\dagger_n\sigma_m,
\end{equation}
which results in coherent exchange of excitations reducing the peak emission rate at small spacings, where the coherent exchange rate becomes large.

The coherent and dissipative dipole-dipole couplings between emitters $n$ and $m$ read
\begin{align} \label{3d-dipole}
   J_{nm} - \frac{i\Gamma_{nm}}{2} = -\frac{3 \pi \Gamma}{\omega_0} \mathbf{d}^\dagger \cdot \mathbf{G}(\mathbf{r}_{nm},\omega_0) \cdot \mathbf{d},
\end{align}
where $\mathbf{d}$ is the transition dipole moment (a complex number in general) and $\mathbf{r}_{nm} = \mathbf{r}_n-\mathbf{r}_m$ is the connecting vector between emitters $n$ and $m$. The Green's tensor $\mathbf{G}(\mathbf{r}_{nm},\omega_0)$ is the propagator of the electromagnetic
field between emitter positions $\boldsymbol{r}_n$ and $\boldsymbol{r}_m$, and for a 3D photonic environment reads
\begin{align}
\mathbf{G}(\mathbf{r}_{nm},\omega_0) = \frac{e^{i k_0 r_{nm}}}{4\pi k_0^2 r_{nm}^3} \bigg[ \left( k_0^2 r_{nm}^2 + ik_0 r_{nm} -1 \right) \mathbb{1}  +  \left(-k_0^2 r_{nm}^2 - 3i k_0 r_{nm} + 3 \right) \frac{\mathbf{r}_{nm} \otimes \mathbf{r}_{nm}}{r_{nm}^2} \bigg],
\end{align}
with $r_{nm} = |\mathbf{r}_{nm}|$ and $k_0 = 2\pi/\lambda_0$, where $\lambda_0$ is the wavelength of light emitted by the emitters. 

For 1D photonic environments such as for two-level emitters coupled via a single-mode waveguide reservoir, the interactions are given by~\cite{sheremet2023waveguide}
\begin{equation} \label{supp:wg}
       J_{nm} - \frac{i\Gamma_{nm}}{2} =  -\frac{i\Gamma}{2}\ \mathrm{exp}\Big(i k_\mathrm{0} |{x_n}-x_m| \Big),
\end{equation}
which exhibits an infinite range, periodic modulation with $k_{0}=\omega_0/c$ being the wavevector of the waveguide mode on resonance with the transition frequency, and it is assumed that the emitters are positioned at $\{x_n\}$ along the waveguide.

\end{document}